\documentclass[aps,prx,twocolumn,superscriptaddress]{revtex4-1}

\usepackage{graphicx}
\usepackage{dcolumn}
\usepackage{amsfonts,amsmath,amssymb,bm}
\usepackage{subcaption}
\usepackage{xcolor}
\usepackage[utf8]{inputenc}
\usepackage[colorlinks=true,allcolors=blue]{hyperref}

\begin{document}

\title{Theoretical study of quantum emitters in two-dimensional silicon carbide monolayer}
\date{\today}
\author{Q. Hassanzada}
\affiliation{Department of Physics, Isfahan University of Technology, Isfahan, 84156-83111, Iran}

\author{I. Abdolhosseini~Sarsari}
\affiliation{Department of Physics, Isfahan University of Technology, Isfahan, 84156-83111, Iran}

\author{A. Hashemi}
\affiliation{Department of Applied Physics, Aalto University, P.O. Box 11100, 00076 Aalto, Finland}

\author{A. Ghojavand}
\affiliation{Department of Physics, Isfahan University of Technology, Isfahan, 84156-83111, Iran}

\author{A. Gali}
\affiliation{Department of Atomic Physics, Budapest University of Technology and Economics, Budafoki út 8, 1111 Budapest, Hungary}

\author{M. Abdi}
\affiliation{Department of Physics, Isfahan University of Technology, Isfahan, 84156-83111, Iran}

\begin{abstract}
The features of some potential single-photon sources in two-dimensional silicon carbide monolayers is studied via ab-initio calculation and group theory analysis.
A few point defects in three charge states (negative, positive and neutral) are considered.
By applying performance criteria, Stone-Wales defects without and with combination of antisite defects are studied in details.
The formation energy calculations reveal that neutral and positive charge states of these defects are stable.
We compute the zero-phonon-line energy, the Huang-Rhys (HR) factor and the photoluminescence spectrum for the available transitions in different charge states.
The calculated HR values and the related Debye-Waller factors guarantee that the Stone-Wales defects have a high potential of performing as a promising single-photon emitter.
 
\end{abstract}

\maketitle
\newcommand{\RNum}[1]{\uppercase\expandafter{\romannumeral #1\relax}}

\section{INTRODUCTION}
Two-dimensional (2D) materials have recently received much attraction in the scientific communities because of their unique properties and extensive applications.
A prominent aspect of 2D materials is their applications in quantum technologies where single photons are essential.
Observation and investigation of single photons in 2D materials such as $h$-BN, transition metal dichalcogenides, WO$_{3}$, and GaSe
\cite{Toth2019, Tran2017, Palacios-Berraquero2017, He2015, Tonndorf2017, Branny2016, Chakraborty2016},
has raised a great interest toward this class of materials in the field
of single-photon emitters (SPE). SPE's in 2D materials are applicable in many quantum technologies ranging from quantum nanophotonics to quantum sensing and quantum information processing~\cite{Aharonovich2017}.
In such materials, the 2D nature of 
the host considerably improves the photon-extraction efficiency, provides more control over the defect implantation techniques, eases coupling to the waveguides, and provides compatibility with other 2D materials~\cite{Sajid2019}.
These advantages inspire researchers to search and find 2D materials in which all the desired properties of an efficient single-photon source are met. Over the past few years, group IV semiconductors have shown significant potential to host SPE's,
in such a way that most ideal single-photon sources 
were investigated just in diamond and silicon carbide (SiC) 
\cite{Aharonovich_2011,Lohrmann_2017}.
Among these traditional three-dimensional (3D) bulk materials, SiC not only emits room temperature single photons but also provides the brightest nonclassical light ever
\cite{Castelletto2013,Khramtsov2018}. 
The advantages of 3D SiC in hosting SPE's makes one consider the investigation of the local point defects in its 2D form as potential color centers.

Even though the fabrication of 2D-SiC is still challenging~\cite{Lin2012, Szabo2009, Chabi_2016,Lin2015,Susi2017},
the recent observation of SiC nano-grain assembly in graphene oxide pores~\cite{Susi2017} and investigation of quasi two dimensional SiC~\cite{Lin2015} affirm that the fabrication of two dimensional SiC is about to emerge.
Besides, theoretical calculations reveal that many possible 2D structures of SiC are probable in which Si$_{0.5}$C$_{0.5}$ structure is the most stable one~\cite{Shi2015}. Despite the preference of $sp^{3}$ hybridization of the Si atoms, the Si$_{0.5}$C$_{0.5}$ structure, which here on we call it 2D-SiC, is predicted to be flat with $sp^{2}$ hybrid bonds. The band gap of this structure is larger than 3~eV making it an appropriate host for color centers.
Considering the high potential of bulk SiC in producing color centers, the predicted wide band gap of 2D-SiC 
and the superiority of 2D-materials in some applications, defects in two-dimensional SiC might be capable of emitting ideal single photons.

In this work, we present a study on electronic and optical properties of some favorable defects in 2D-SiC utilizing first principle calculations assisted by group theoretical analysis to explore SPEs.
The study includes vacancy, antisite, substitutional and Stone-Wales (SW) defects.
To investigate potential quantum emission defects, we first calculate zero-phonon-line (ZPL) energy for all possible transitions.
Afterwards, we evaluate the Huang-Rhys (HR) factor and the related photoluminescence (PL) spectrum for the most likely cases.
Our first-principles study reveals that the family of SW defects---SW (a simple $90^{\circ}$ rotation of a silicon-carbon pair), SW-Si$_{\text{C}}$ (a silicon-carbon pair rotation with silicon instead of carbon), SW-C$_{\text{Si}}$ (a silicon-carbon pair rotation with carbon instead of silicon)---have a high potential for emitting high-quality single photons.

\section{METHOD}
To analyze point defects embedded in 2D-SiC, density functional theory (DFT) is used as 
implemented in the \textsc{vasp} code
\cite{Kresse1996}.
The electronic structure is calculated using the generalized gradient approximation by Perdew, Burke and Ernzerhof (PBE) 
\cite{Perdew1996}. 
Since isolated defects embedded in a wide band gap material cause isolated states inside the band gap, the initial step is to calculate the band gap correctly. Due to the
well-known failure of DFT in the calculation of the band gap, hybrid exchange functional of Heyd, Scuseria, and Ernzerhof (HSE06) is applied
\cite{Heyd2003,Krukau2006}
to overcome this problem. HSE06 hybrid functional calculates the band gap and the charge transition levels of group-IV semiconductors within 0.1 eV accuracy 
\cite{Peter2010}.
The excited-state calculations are carried out using constrained-occupation DFT (CDFT) approach which gives reliable results 
for the ZPL energy when compared to experimental data
\cite{Thiering2018}.
Because of the limited computational resources at hand, a supercell of 72 atoms ($6\times 6$) is chosen to conduct preliminary research on the electronic structure of different defects. When a few candidates are handpicked, the supercell size is magnified to 128 atoms 
($8\times 8$) to study the convergence of the calculations. For $6\times 6$ supercell, a vacuum of 12 ${\AA}$ and for $8\times 8$ supercell
a vacuum of 24 ${\AA}$ are considered to separate the periodic images. A single k-point ($\Gamma$ point) is applied to sample the Brillouin zone. While a single k-point might not be numerically convergent for a supercell of 72 atoms, but the obtained results serve as a guidance to pick up the most likely candidates for further investigations.
An energy cutoff of 450 eV is utilized for the plane-wave basis set within the projector augmented-wave method (PAW)
\cite{Blochl1994,Bengone2000}.
The ground state electronic structure is obtained by performing geometry relaxation in which internal positions of atoms are relaxed until forces on them are smaller than 0.02 eV/\AA. Within the CDFT method, the ions are relaxed to reach the global minimum of adiabatic potential energy surface in the excited state. In this method, one electron is promoted from an occupied Kohn-Sham level to an empty Kohn-Sham level. The ZPL energy is attained by the difference in total energy between two electronic states minima.

The PL spectrum is calculated within the HR theory which for the first time was proposed by Huang and Rhys in 1950 
\cite{Huang1950}. 
One important quantity in HR theory is the partial Huang-Rhys factor. This factor helps us in assessing a point defect as a good single photon emitter. The factor is given by
\begin{equation}
S_{k}=\frac{\omega_{k}q_{k}^{2}}{2\hbar}\text{,}
\end{equation}
where $\omega_{k}$ is frequency of the $k$th phonon mode, $\hbar$ is the reduced Plank constant and $q_{k}$ is defined as
\begin{equation}
q_{k}=\sum_{\alpha, i} m_{\alpha}^{\frac{1}{2}}\big(R^{(e)}_{\alpha, i}-R^{(g)}_{\alpha, i}\big)\Delta r_{k,\alpha, i}\text{,}
\end{equation}
Where, $m_{\alpha}$ is the mass of atom $\alpha$, $R^{(e)}_{\alpha, i}$ and $R^{(g)}_{\alpha, i}$ are equilibrium atomic coordinate of atom $\alpha$ along the direction $i$ in the excited state and the ground state and $\Delta r_{k,\alpha, i}$ is the normalized displacement vector of atom $\alpha$ along the direction $i$ in phonon mode $k$. Sum of the partial HR factor over all phonon modes $k$ gives the total HR factor. This quantity determines coupling between electronic and vibrational states and characterizes the phonon side band.
The PL intensity is defined as:
\begin{equation}
L(\hbar w)=Cw^{3}A(\hbar w)
\end{equation}
where $C$ is the normalization constant and $A(\hbar w)$ is the optical spectral function given by
\begin{equation}
A(\hbar w)=\int_{-\infty}^{\infty}G(t)e^{iwt-\gamma\left |t  \right |} dt.
\end{equation}
Here, $\gamma$ is width of the ZPL, $G(t)$ is the generating function and is defined as
\begin{equation}
G(t)=e^{S(t)-S_{\rm tot}},
\end{equation}
where $S(t)$ is the Fourier transform of the partial HR factor, defined as
\begin{equation}
S(\hbar w)=\sum_{k}S_{k}\delta (\hbar w-\hbar w_{k}),
\end{equation}
and $S_{\rm tot}$ is the total HR factor.

\begin{figure*}[t]
\centering
\includegraphics*[scale=0.075]{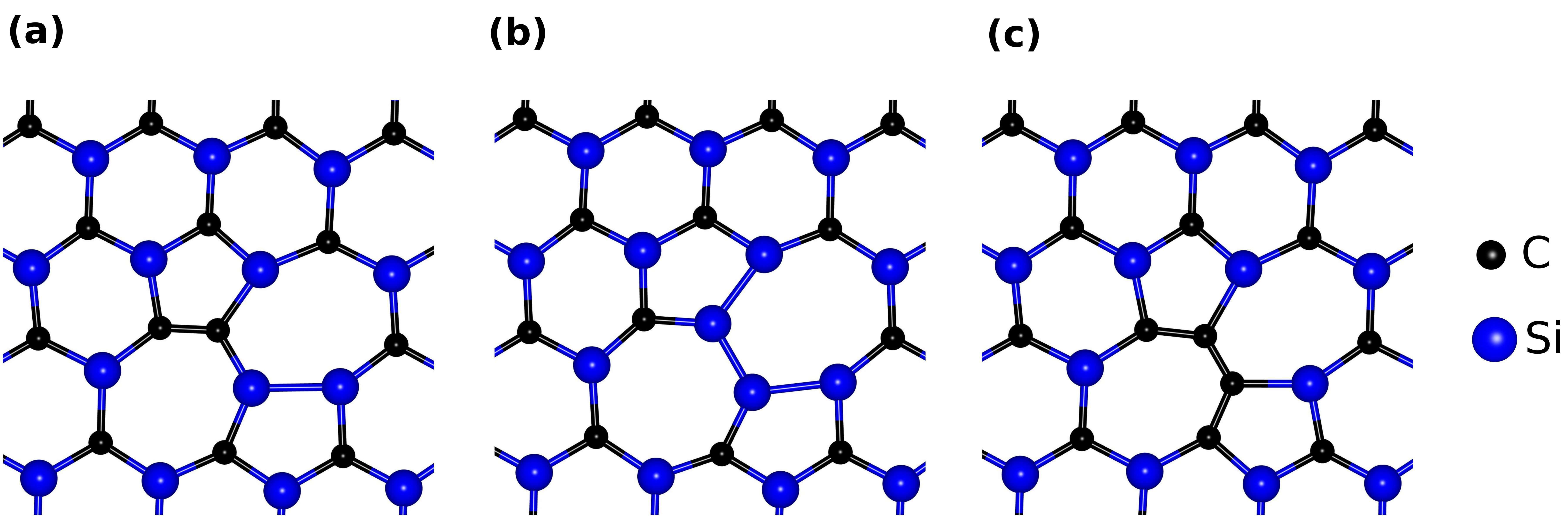}
\caption{Atomic structure of (a) SW , (b) SW-Si$_{\text{C}}$ and (C) SW-C$_{\text{Si}}$ defects. 
Silicon atoms and carbon atoms are displayed with blue and black colors.}
\label{Structure}
\vspace{-3mm}
\end{figure*}

Here the PL spectrum and the HR factor are calculated based on Refs.~\cite{Alkauskas_2014,Gali2016}.
Since PBE functional provides the PL spectrum in good agreement with HSE06 hybrid functional, by applying this functional the geometry
is relaxed until all forces are smaller than $10^{-3}$ eV/\AA, then density functional perturbation theory (DFPT) within the \textsc{vasp} code is used to obtain the phonon eigenfrequencies and eigenvectors. 
A fast, yet efficient, alternative method for determining the quality of the quantum emissions from the defects is the $\Delta Q$ parameter. This parameter specifies the atomic structure change during the excitation. To calculate it we apply one-dimensional configuration coordinate formulation which is represented by the formula
\begin{equation}
\Delta Q^{2}=\sum_{\alpha, i} m_{\alpha } \Delta R_{\alpha, i}^{2}, 
\end{equation}
where $\Delta R_{\alpha, i}=R^{(e)}_{\alpha,i} -R^{(g)}_{\alpha,i}$.
Even though not exactly, the high (low) values of $\Delta Q$ are proportional to high (low) values of the HR factor. Since it is not computationally expensive to calculate $\Delta Q$, we use it as a filtering parameter for thorough study of a few defects.
The stability of charged defects is determined by evaluating the formation energy as a function of the Fermi level as

\begin{align}
 E_{q}^{\rm F}(\epsilon _{f}) = &E_{\text{SiC},q}^{\rm tot}-E_{\text{SiC},p}+q\big[\epsilon_{\rm vbm}^{\rm pri} +\epsilon_{\rm F} -\Delta V \big] \nonumber\\
& -n_{x}\mu _{x}+E_{q}^{\rm corr}.
\end{align}
In the above equation, $E_{q}^{\rm tot}$ is the total energy of 2D-SiC supercell containing the charged defects, $E_{\text{SiC},p}$ and $\epsilon_{\rm vbm}^{\rm pri}$ are the total energy of a pristine 2D-SiC and pristine valence band maximum, $\mu_{x}$ is the chemical potential of atoms and $n_{x}$ determines the number of atom of type $x$ added (positive) or removed (negative) from the pristine supercell and $E_{q}^{\rm corr}$ is the finite-size electrostatic correction. Here for calculating 
the correction term the \textsc{CoFFEE} code is utilized~\cite{Naik2018}.
For the growth condition, it is assumed that in the Si-rich condition, 2D-SiC is in equilibrium with bulk Si whereas in the C-rich condition it is in equilibrium with graphite.

\subsection*{Group theoretical analysis}
In order to assist the DFT calculations, we invoke group theory analysis for further understanding of the electronic and spin properties of the studied point defects.
As it will become clear shortly, the focus of our study is put on the Stone-Wales defect varieties. The molecular symmetry of these defects is relatively low as they assume either a $C_{2v}$- or $C_s$-point group symmetry.
The molecular orbitals (MOs) that are a linear combination of the atomic orbitals are form by adapting the symmetry of the molecules. These MOs provide a basis that diagonalizes the configuration Hamiltonian.
Each orbital has the symmetry of one irreducible representation (irr) of the point group.
By investigating the character tables and comparing the MOs provided by the ab-initio calculations, one identifies their irreducible representation.
The multi-electron states are then obtained by first determining the orbital occupations, the spin state of the collective electrons, and finally the tensor product of the spatial wavefunction with the spin part.
Here, the neutral charge defects can only assume singlet in their ground state and both singlet and triplet states in the excited state and all other combinations are rejected by Pauli's exclusion principle. Meanwhile, the positively charged defects are only spin-doublet.
\begin{figure*}[tb]
\includegraphics*[width=0.95\textwidth]{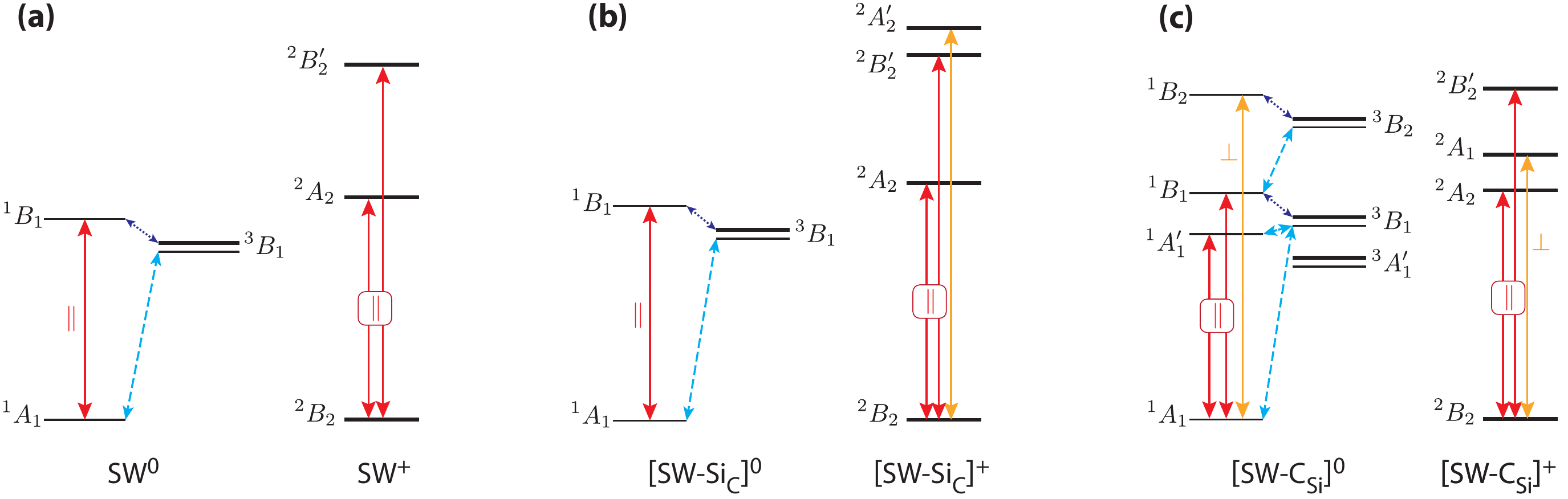}
\caption{\label{fig:jablonski}%
The electronic structure (to the scale) of the three SW defects thoroughly studied in this work in the neutral and positive charge states. The solid arrows indicate allowed dipole transitions. The transitions with in-plane polarization are expected to give brighter emissions (red) while the transitions induced by the out-of-plane dipole moments are presented in orange (weaker excitation).
Dashed and dotted blue arrows show the non-radiative transitions that can be induced by spin interactions.}
\end{figure*}
\begin{table}[!ht]
\centering
\setlength\extrarowheight{-30pt}
\setlength\tabcolsep{5pt}
\caption{\label{transitions}
The calculated ZPL energy and $\Delta Q$ value for defects possible transitions using HSE06 hybrid functional in $6\times 6$ supercell. The superscripts ``up" and ``down" refer to spin up and spin down channels.  
}
\begin{tabular}{ c | c | c | c }

defect &  $E_{\rm ZPL}^{\rm up}$(eV) &  $E_{\rm ZPL}^{\rm down}$(eV) & $\Delta Q$(\AA$\cdot$amu$^{1/2}$) \\
\hline\hline
              &      &      &    \\
V$_{\text{C}}^{+}$   & 2.59 &  ---    & 2.41   \\
V$_{\text{C}}^{-}$   &  ---    & 0.59 & 1.23   \\
              &      &      &        \\
              \hline
              &      &      &        \\
V$_{\text{Si}}$-V$_{\text{C}}$      & 0.917&  ---    & 1.25  \\
                    & 1.76 &  ---    & 2.16  \\
V$_{\text{Si}}$-V$_{\text{C}}^{+}$  & 1.02 &  ---    & 1.56  \\
                    & 2.21 &  ---    & 1.20  \\
V$_{\text{Si}}$-V$_{\text{C}}^{-}$  &   ---   & 1.51 & 1.08  \\
                    &      &      &       \\
                    \hline
                    &      &      &        \\
V$_{\text{C}}$-C$_{\text{Si}}$      & 1.48 & ---     & 0.83  \\            
                    & 1.05 &  ---    & 0.85 \\
V$_{\text{C}}$-C$_{\text{Si}}^+$  & 0.86 &   ---   & 1.25 \\
                    & 1.69 & ---     & 0.88 \\
                    & 2.32 & ---     & 0.93 \\
V$_{\text{C}}$-C$_{\text{Si}}^-$  &  ---    & 2.18 & 0.63 \\
                    & ---     & 0.41 & 0.87 \\
                    &      &      &      \\
                    \hline
                    &      &      &        \\
V$_{\text{Si}}$      & ---    & 1.99 & 1.03 \\
              & ---    & 1.05 & 0.94 \\
V$_{\text{Si}}^{+}$  & ---    & 0.88 & 0.58 \\
              & ---    & 1.90 & 1.05 \\
              & ---    & 1.97 & 1.11 \\
V$_{\text{Si}}^{-}$  & ---    & 2.03 & 1.42 \\
              &     &      &      \\
              \hline
              &      &      &        \\
SW          & 1.47 & ---     & 0.46   \\
SW$^{+}$      & 1.63 &  ---    & 0.49   \\
SW$^{-}$      &  ---    & 1.60 & 0.53   \\
              &      &      &        \\
              \hline
              &      &      &        \\
SW-Si$_{\text{C}}$        & 1.54 &  ---    & 0.39 \\
SW-Si$_{\text{C}}^+$    & 1.77 &  ---    & 0.40 \\
                   & 2.64 &  ---    & 0.46 \\
SW-Si$_{\text{C}}^-$    & ---     & 3.35 & 0.35 \\
                   & ---     & 3.29 & 0.22 \\
                   & ---     & 1.50 & 0.42 \\
                   &      &      &      \\
                   \hline
                   &      &      &        \\
SW-C$_{\text{Si}}$        & 1.65 &  ---    & 0.53 \\
                   & 1.32 &  ---    & 2.32 \\
                   & 2.28 &  ---    & 0.63 \\
SW-C$_{\text{Si}}^+$    & 1.73 &  ---    & 0.57 \\
                   & 1.33 &  ---    & 2.51 \\
                   & 2.51 &  ---    & 0.81 \\
SW-C$_{\text{Si}}^-$    &  ---    & 1.67 & 0.48 \\
                   &      &      &       \\
\hline\hline
\end{tabular}
\end{table}

The group theory allows us to predict potentially nonzero matrix elements of the Hamiltonians that are perturbatively included in the analysis. The spin-orbit interaction is one of the crucial interactions that yet remains a perturbation in the system as SiC is composed of relatively light atom species. The interaction induces non-radiative transitions between the multi-electron states. Another critical effect which lies at the heart of our analysis is the electric dipole interaction that determines dark and bright states. Those states that their dipole moment can couple to the ground state via the external electric field, provided by electromagnetic radiation determine the selection rules~\cite{Abdi2018, Li2020}.
In this part of study, the value of matrix elements $\langle\psi|O|\phi\rangle$ is vanishing if their total irreducible symmetry is not that of totally symmetric irreducible representation, here $A_1$ and $A'$ for the $C_{2v}$ and $C_s$, respectively. In the mathematical expression $\Gamma(\psi)\otimes \Gamma(O) \otimes \Gamma(\phi) \not\supset \Gamma_1$, where $\Gamma(X)$ is the irreducible representation of the object $X$ and $\Gamma_1$ is the totally symmetric irr.
The selection rules are obtained by studying the electric dipole interaction $H_{\rm dp} = -\mathbf{d}\cdot\mathbf{E}$, where $\mathbf{d}=-e(x,y,z)$ is the dipole moment of the electronic state and $\mathbf{E}$ is the electric field vector. $\mathbf{d}$ transforms like a polar vector, and its components have a different irr as listed in the character tables. The polarization of radiative transitions is thus predicted by inspecting those components with non-vanishing matrix elements.
\begin{figure}[b]
\centering
\includegraphics*[scale=0.45]{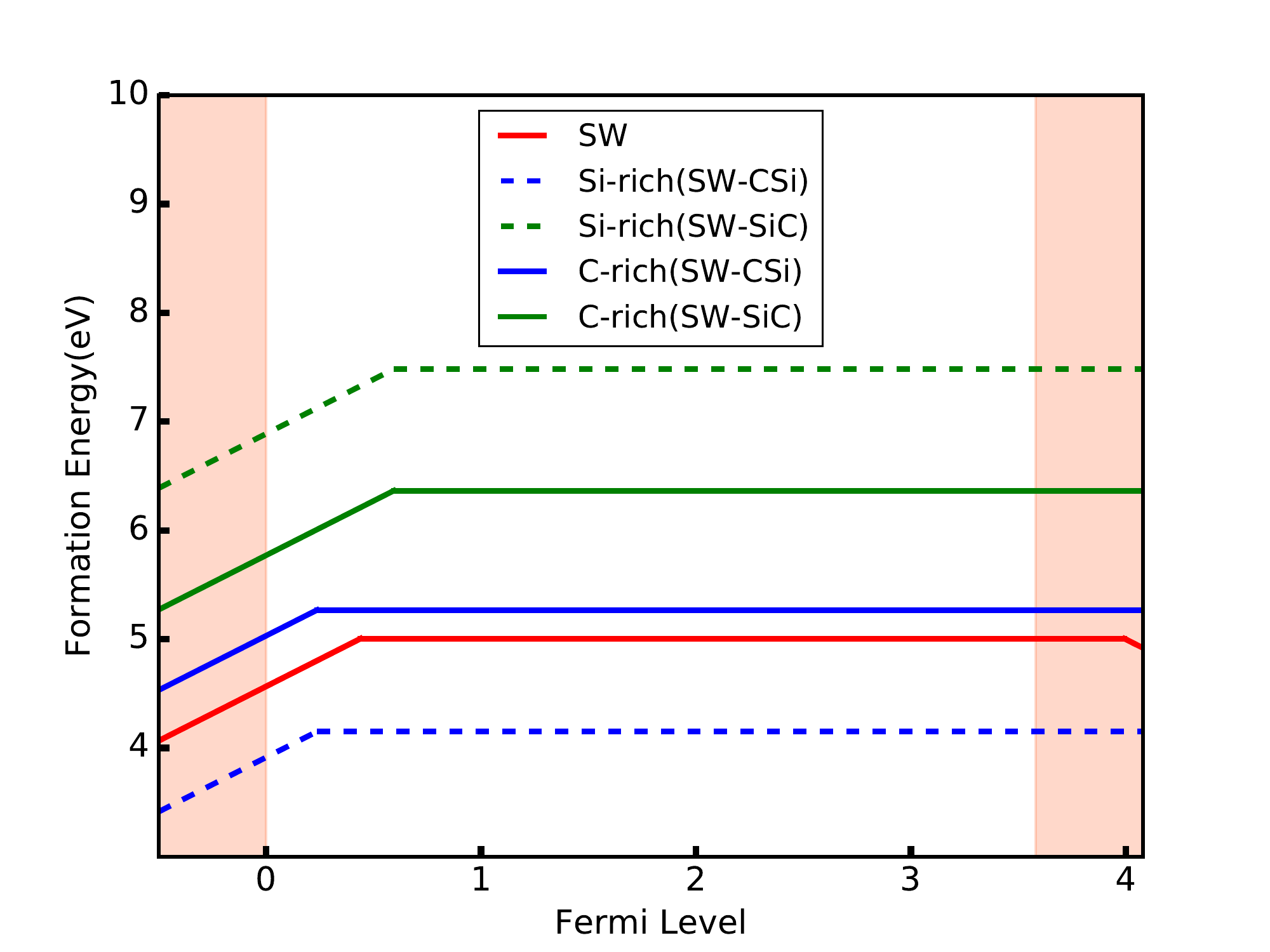}
\caption{\label{formation} Formation energy as a function of the Fermi level for SW (red lines), SW-Si$_{\text{C}}$ (green lines) and SW-C$_{\text{Si}}$ (blue lines) defects in C-rich (solid lines) and Si-rich (dashed lines) growth conditions. Both growth conditions for SW defect are the same.}
\end{figure}
\begin{figure*} [t]
\centering
\includegraphics*[scale=0.05]{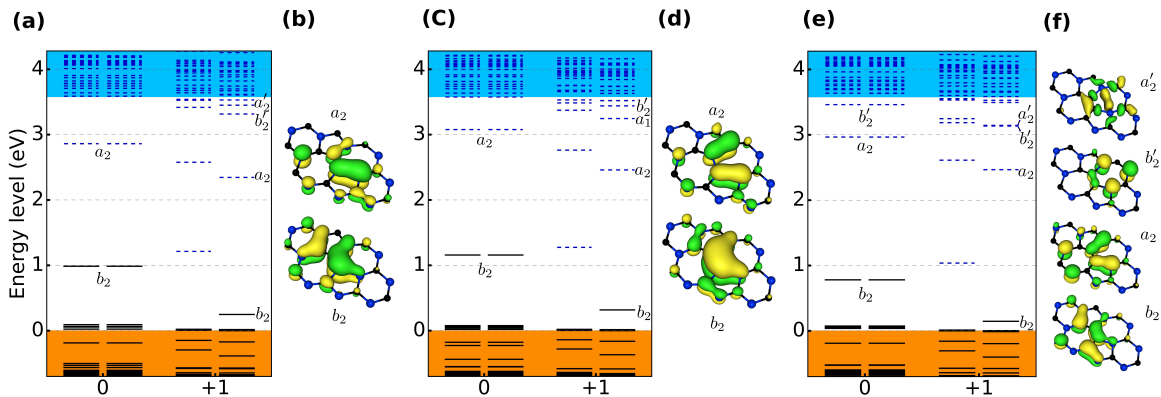}
\vspace*{0mm}
\caption{Kohn-Sham ground state electronic levels of (a) SW , (c) SW-Si$_{\text{C}}$ and (e) SW-C$_{\text{Si}}$ defects. 
Occupied and unoccupied levels are shown with solid black lines and blue dashed lines. Charge density plots for neutral defect states are displayed in 
(b) SW, (d) SW-Si$_{\text{C}}$, (f) SW-C$_{\text{Si}}$.
The symmetry of molecular orbitals are shown both next to the energy levels and the charge density plots. The $C_{2v}$ symmetry labels are also used for the SW defect.}
\label{Electronic-structure}
\end{figure*}
\section{RESULTS AND DISCUSIONS}
In our DFT calculation, we find that 2D-SiC remains in a plane which agrees with previous works\cite{Susi2017,Shi2015}. The Si-C bond distance is calculated to be 1.78 \AA\ with PBE functional and 1.77 \AA\ with HSE06 hybrid functional. We evaluate a large band gap of 3.58 eV for 2D-SiC with HSE06 hybrid functional which is larger than any main polytypes 
of SiC.
We survey various defects which are probable in 2D-SiC. These defects include Si/C vacancy, vacancy-antisite, divacancy, Si/C substitutional,
Stone-Wales defects (SW, SW-Si$_{\text{C}}$, SW-C$_{\text{Si}}$). Stone-wales defects involve an in-plane $90^{\circ}$ rotation of a bond in $sp^{2}$-bonded materials. They already have been investigated in graphene and silicene
\cite{Lin2012,Chabi_2016,Lin2015,Susi2017}. Due to this fact, their formation in 2D-SiC are predictable. We study defects electronic structure in $6\times 6$ supercell, inspecting their properties as a color center candidate. By looking for the electronic levels inside band gap and the possible transitions between these levels for every defect, we find that both of Si$_{\text{C}}$ (silicon instead of carbon) and C$_{\text{Si}}$ (carbon instead of silicon) defects are inappropriate as they do not introduce any states within the band gap. On the other hand, the possible transitions of the neutral V$_{\text{C}}$ defect exhibit small values of ZPL energy, putting them in the infrared region. We, therefore, exclude these defects from our further analysis. For all surveyed defects, we calculate the possible excited states within the CDFT approach. The results are summarized in Table~\ref{transitions}.
We employ the $\Delta Q$ parameter to handpick the defects so that those defects with $\Delta Q$ parameters smaller than 0.6 \AA$\cdot$amu$^{1/2}$ are considered for the detailed study. One clearly notices that Stone-Wales defects SW, SW-Si$_{\text{C}}$, SW-C$_{\text{Si}}$ and V$_{\text{Si}}^{+}$ satisfy this criterion. Even though it suggests that they could have the potential to emit single photons, further study is needed to confirm this speculation. Our focus is set on optical photon emitters. Thus, due to the small value of the ZPL energy for V$_{\text{Si}}^{+}$ defect our investigation becomes restricted to the Stone-Wales defects.
The atomic structure of these defects is shown in Fig.~\ref{Structure}.
We investigate the relative stability of these defects by computing the formation energy in neutral, negative, and positive charge states in $8\times 8$ supercell. The results for HSE06 hybrid functional are presented in Fig.~\ref{formation}. The obtained formation energy shows that the negative charge state is not stable for any of the Stone-Wales defects as the (0/$-$) charge transition levels for SW, SW-Si$_{\text{C}}$ and SW-C$_{\text{Si}}$ defects are at EV+4.00 eV, EV+4.30 eV and EV+4.19 eV, respectively, where EV is the valence band maximum. 
Unlike the negative charge states, the neutral charge states are stable as the (+/0) charge transition levels are at EV+0.43 eV, EV+0.59 eV and EV+0.23 eV, respectively. We also consider the positive charge state of these defects stable because their highest occupied molecular orbital (HOMO) lies between the (+/0) charge transition level and the valence band maximum within the band gap.

Before going through the detailed ab-initio calculation on these defects, we provide a group of theoretical calculations with an emphasize on the electronic structure and the possible transitions, both radiative and non-radiative. The substitutional SW defects have a $C_{2v}$ symmetry and irr of their orbitals are as assigned in Fig.~\ref{Electronic-structure}. 
The neutral charge defects have a fully occupied singlet orbital in their ground state. Thus, we expect a spin-singlet state for these cases. In the excited states, however, both singlet and triplet spin states are allowed that divides the radiative transition channels into triplet and singlet as the dipole transitions are spin state preserving in the zeroth order of spin-orbit slater states.
This allows us to construct the electronic structure of the few lowest states and determine the polarization of the radiative transitions according to the matrix element symmetries [see Fig.~\ref{fig:jablonski} for the structure as well as the transitions].
The positively charged defects provide only one electron in the band gap, which implies the one have to expect spin-doublet states. We find that the order of orbitals is slightly shuffled by subtraction of one electron from the defect. The electronic structure, thus, behaves differently.
Our analysis on the spin-orbit interaction suggests that non-radiative transition through the shelving states is also possible.
The pure SW defect has a reduced symmetry of $C_s$, which results in an increased dipole transition probability. For this reason, we expect brighter emissions. However, note that  for the sake of clarity we have assigned $C_{2v}$ irr labels to the MOs and the multi-electron states in Fig.~\ref{fig:jablonski} as well as Fig.~\ref{Electronic-structure}.

We recalculate the electronic structure of these defects in neutral and positive charge states in $8\times 8$ supercell. Their electronic structure is shown in Fig.~\ref{Electronic-structure}. If we take a look at this structure, we will find that all the neutral charge states and all the positive charge states are similar. It is because carbon and silicon belong to the same group in the periodic table of the elements as it was demonstrated in group-IV vacancy color centers in diamond \cite{Thiering2018}. Localization of charge density plot is also displayed in Fig.~\ref{Electronic-structure}. It determines electronic defect states in the band gap.
\begin{table*}[t]
\centering
\setlength\extrarowheight{5pt}
\setlength\tabcolsep{5pt}
\caption{\label{ZPL-transitions}
The ZPL energy , Huang-Rhys and Debye-Waller factors of Stone-Wales defects for the first, second and third excitations in $8\times 8$ supercell.}
\begin{tabular}{ c | c | c | c | c|c|c|c|c|c}
defect & First (eV) & HR & DW &  Second (eV) & HR & DW&  Third (eV) & HR& DW \\
\hline\hline
SW$^{0}$      &  {\textcolor{red}{1.40}}    &  {\textcolor{red}{0.94}}    &  {\textcolor{red}{39\%}} &   ---   &  ---  &  ---    &--- &--- &---\\
SW$^{+}$      &  1.60    &   1.23  & 29\% &   2.65   &   2.16  &  12\%    & 3.26 & 2.43 & 9\% \\

SW-Si$_{\text{C}}^{0}$     &  {\textcolor{red}{1.50}}    &  {\textcolor{red}{0.74}}    &  {\textcolor{red}{48\%}  } &   ---  &  ---  &  ---  & --- & --- &---\\
SW-Si$_{\text{C}}^{+}$     & 1.71   &  0.93   &  39\% &   2.61   &  1.10  &  33\%   & 2.64 & 11.23 & 0\%\\

SW-C$_{\text{Si}}^{0}$    &  {\textcolor{red}{1.62}}    &   {\textcolor{red}{1.43}}   &  {\textcolor{red}{24\%} } &  1.45   & 26.83 &  0\%  & 2.28  & 2.39  &9\%\\
SW-C$_{\text{Si}}^{+}$    &   1.72   &   1.71   &  18\%   &  1.80    & 9.50  &  0\%   &  2.56 &3.32&4\%\\
\hline\hline
\end{tabular}
\end{table*}
\begin{figure*}[ht!]
\centering
\includegraphics*[scale=0.43]{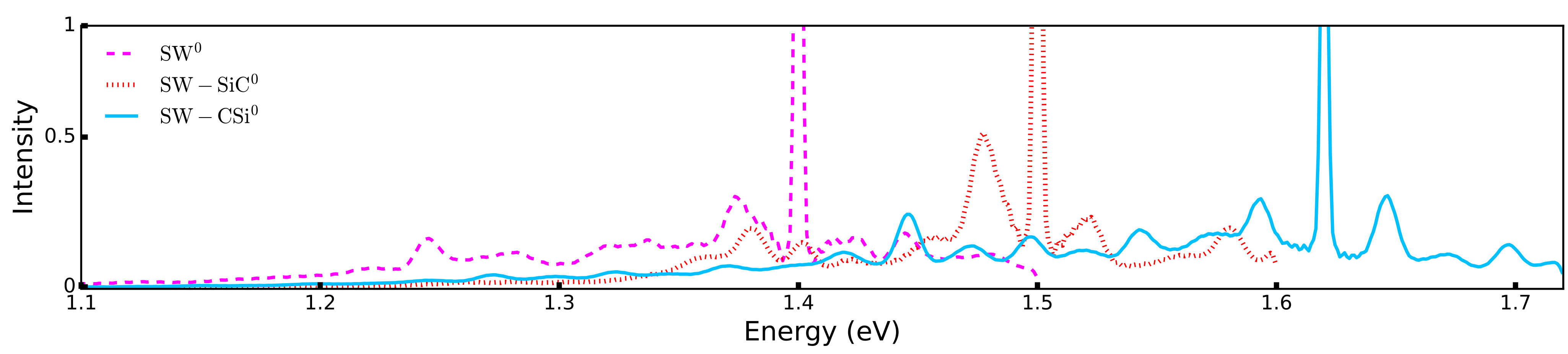}
\caption{\label{PL}%
Photoluminescence line shape of the first excitation of neutral Stone-Wales defects.}
\end{figure*}In this massive supercell, we inspected the defects viability in emitting single photons by evaluating the ZPL energy and the HR factor for every electronic transition. These values were displayed in Table~\ref{ZPL-transitions}. In the table, the Debye-waller (DW) factor describes the ratio of the emission of coherent photons vs. total number of emitted photons including the phonon assisted ones which compromise the coherence. This factor is related to the HR factor and determines the weight of the ZPL. The values of DW factor are up to 50 percent in 2D-SiC monolayer. It confirms the high-quantity of these defects in producing single photons. We should mention that these values would be considerably higher in multilayer of 2D-SiC because the Van der Waals forces between layers decrease the structural changes during the excitation, which leads to higher amounts of DW factor. This difference was also observed in hexagonal boron nitride \cite{Tran2015}.
In Table~\ref{ZPL-transitions} the ZPL energies range from 1.4 to 3.3 eV. It demonstrates that observation of single photons would be possible in visible and near-infrared regions which makes them promising candidates with quantum 
functionalities in quantum communication, quantum information processing and biological sensing. For every defect, we also plotted the PL spectrum of the transition with the highest value of the HR factor, which is shown by red color in Fig.~\ref{PL}. The PL spectrum provides good corroborations and supports the future experimental investigation of single-photon sources which is related to Stone-Wales defects in 2D-SiC. 

\section{CONCLUSION and outlook}
Our first-principles calculation has led us to introduce the most probable single-photon emitter candidates in two dimensional SiC monolayer. We calculated the ZPL energy, the Huang-Rhys factor and the photoluminescence spectrum for some selected defects. The values of the HR factor are in the range of 0.74 to 3.32 (Debye-Waller factor of $48\%$ to $4\%$). Our ab-initio analysis demonstrated that Stone-Wales defects (SW, SW-Si$_{\text{C}}$, SW-C$_{\text{Si}}$) in neutral and positive charge states are capable of emitting single photons.
The group theory analysis for the SW neutral and charged defects predicted the polarization of radiative and non-radiative transitions, respectively. Increasing dipole transition probability in pure SW defect with lower symmetry convinced us of having brighter emissions in these defects. Our analysis on the spin-orbit interaction suggests that non-radiative transition through the shelving states is also possible.
The plotted PL spectra of the transitions with the highest value of the HR factor suggest the following opportunities for future experimental investigation of single-photon sources which are related to the Stone-Wales defects in 2D-SiC.

In this article, a novel two-dimensional material is proposed as a host for single-photon sources and actively encourages other scientific groups who are interested in SPE's field to endeavour for the fabrication of 2D-SiC and explore other fascinating properties of this material. We find that typical defects in 2D materials, Stone-Wales defects, are promising quantum emitters in 2D-SiC which may direct researchers to seek other bright emitters in this material. We note that the presence of metastable triplet states may be harnessed as quantum bits in the neutral Stone-Wales defects but further investigations are required to study this phenomenon in Stone-Wales defects and other potential quantum bits in 2D-SiC.

\section{ACKNOWLEDGMENTS}
The authors gratefully acknowledge 
the Sheikh Bahaei National High Performance Computing Center (SBNHPCC) for providing computing facilities
and time. SBNHPCC is supported by scientific and technological department of presidential office and 
Isfahan University of Technology (IUT).
We gratefully acknowledge the help provided by Gerg\H{o} Thiering. A.G.\ acknowledges the National Quantum Technology Program and Natinonal Excellence Program (project numbers 2017-1.2.1-NKP-2017-00001 and KKP129866).
MA acknowledges support by INSF (Grant No. 98005028).
\bibliography{2dSiC}
\end{document}